\documentclass[10pt,twocolumn]{article}
\usepackage{epsf,latexsym}

\newcommand{\sect}[1]{\setcounter{equation}{0}\section{#1}}

\newcommand{\ket}[1]{\left|\, #1\, \right \rangle}

\newcommand{\ra}{{\rightarrow}}

\newcommand{\be}{\begin{equation}}
\newcommand{\ee}{\end{equation}}

\newcommand{\apl}{Appl. Phys. Lett.}

\newcommand{\pr}{Phys. Rev.}

\newcommand{\prl}{Phys. Rev. Lett.}

\def\bbbc{{\mathchoice {\setbox0=\hbox{$\displaystyle\rm C$}\hbox{\hbox
to0pt{\kern0.4\wd0\vrule height0.9\ht0\hss}\box0}}
{\setbox0=\hbox{$\textstyle\rm C$}\hbox{\hbox
to0pt{\kern0.4\wd0\vrule height0.9\ht0\hss}\box0}}
{\setbox0=\hbox{$\scriptstyle\rm C$}\hbox{\hbox
to0pt{\kern0.4\wd0\vrule height0.9\ht0\hss}\box0}}
{\setbox0=\hbox{$\scriptscriptstyle\rm C$}\hbox{\hbox
to0pt{\kern0.4\wd0\vrule height0.9\ht0\hss}\box0}}}}

\def\bbbz{{\mathchoice {\hbox{$\sf\textstyle Z\kern-0.4em Z$}}
{\hbox{$\sf\textstyle Z\kern-0.4em Z$}}
{\hbox{$\sf\scriptstyle Z\kern-0.3em Z$}}
{\hbox{$\sf\scriptscriptstyle Z\kern-0.2em Z$}}}}

\newcommand{\putfig}[2]{$$\leavevmode\hbox{\epsfxsize=#2 cm 
\epsffile{#1.eps}}$$}

\begin{document}

\twocolumn[\hsize\textwidth\columnwidth\hsize\csname
@twocolumnfalse\endcsname

\vspace{-0.7in}


\begin{center}
{\LARGE \bf
Ballistic Single-Electron Quputer}

\vspace{.2in}
{\large {Radu Ionicioiu, Gehan Amaratunga, Florin Udrea}\\
       {\small\it Engineering Department, University of Cambridge}\\
       {\small\it Trumpington Street, Cambridge CB2 1PZ, UK}\\
       {\small\tt email: ri10001@eng.cam.ac.uk}\\}
\end{center} 

\begin{abstract}
We propose a new solid state implementation of a quantum computer (quputer) using ballistic single electrons as {\em flying qubits} in 1D nanowires. We use a single electron pump (SEP) to prepare the initial state and a single electron transistor (SET) to measure the final state. Single qubit gates are implemented using quantum dots as phase shifters and electron waveguide couplers as beam splitters. A Coulomb coupler acts as a 2-qubit gate, using a mutual phase modulation effect. Since the electron phase coherence length in GaAs/AlGaAs heterostructures is of the order of 30$\mu$m, several gates (tens) can be implemented before the system decoheres.

~\\

\end{abstract}

\vskip2pc]

\sect{Introduction}

Quantum computation is a fundamentally different way of thinking about computation and represents a new paradigm for computing science \cite{dd1,dd2}. Using quantum mechanical phenomena to store and manipulate information, we can achieve results impossible to attend by classical means. This includes Shor's super-fast factoring algorithm \cite{shor} and Grover's search algorithm \cite{grover}, to cite only a few results from a rapid expanding field.

Different models for how to build a quantum computer (or {\em quputer}, for short) have been proposed. Among these are ion traps, NMR quantum computation, cavity QED, single photonics. All of these models have strengths and weaknesses, either due to decoherence or because they are not easily scalable (as in the case of the NMR model).

From the point of view of scalability and miniaturisation, a solid state implementation of a quputer has definite advantages over other implementations. Several models for a solid state quantum computer have been proposed so far \cite{kane}-\cite{schon2}. Recently, Nakamura {\em et al} \cite{nakamura} have experimentally realized the control of a qubit using a superconducting Cooper-pair box.

In this article we propose a new model for a quantum computer. The main idea is to use ballistic electrons as {\em flying qubits} in 1D nanowires. We describe how to prepare and measure the qubit and how to implement 1- and 2-qubit gates. Finally, we construct from these gates an electronic version of the optical quantum Fredkin gate.

\section{The Model}

The essential requirement for any model of a quputer is to maintain the coherence of the qubits during the entire period of computation. In order to do this we use single electrons in ballistic regime in 1D nanowires.

The parameter which characterises the coherence of the system is the {\em phase coherence length} $L_\phi$ over which the electrons maintain their phase coherence. At low temperatures (usually in the range of 10 mK) the phase coherence length is of the order of tens of microns. For example, $L_\phi\sim 18\mu$m for gold \cite{mohanty} and $L_\phi\sim 30-40\mu$m for a GaAs/AlGaAs heterostructure \cite{gaas}.

We choose the following states for the qubit:
\[ \ket{0}= vacuum \]
\[ \ket{1}= single\ electron\ state \]

Thus, the presence of an electron is taken as state $\ket{1}$ and its absence as $\ket{0}$. This is analogue to the single photon quantum computer proposed by various authors. One advantage of this choice is that we can also use a {\em dual rail} representation for the qubit \cite{simple_qc}, ie use two physical qubits to represent one logical qubit: $\ket{0}\equiv\ket{10}\ , \ \ket{1}\equiv\ket{01}$, where $\ket{ab}=\ket{a}\ket{b}$ are the two physical qubits.

There are three phases essential to any quantum computation:\\
(i) state preparation;\\
(ii) gating;\\
(iii) state measurement.\\
We now describe each of these in detail.

~\\
{\bf (i) State preparation}\\
We use a single electron pump (SEP) to prepare the initial state for each qubit. A SEP is schematically shown in the figure
\putfig{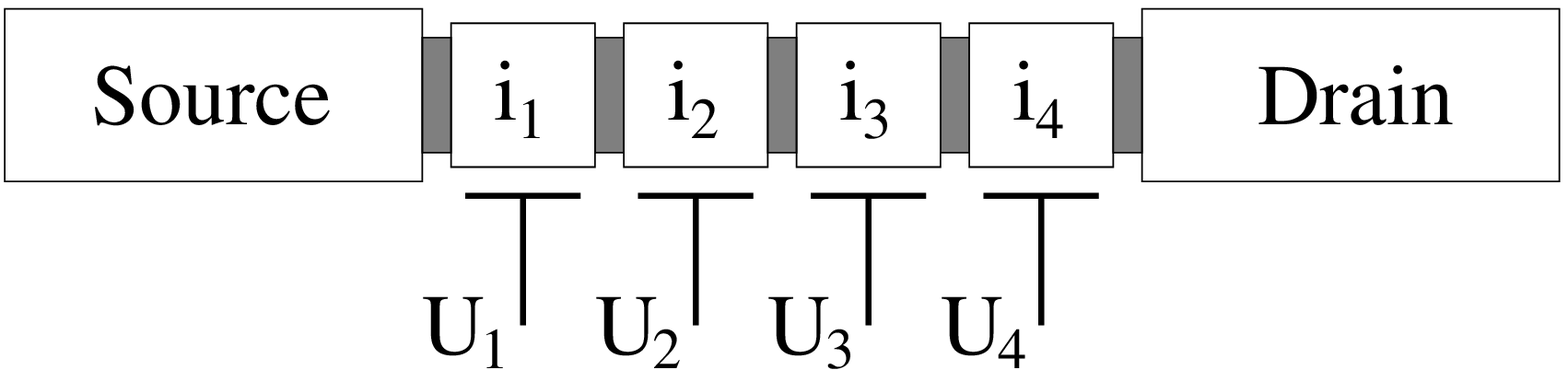}{5}
Between a source and a drain there are several conducting islands separated from each other (and from the source and the drain) by tunnelling barriers. The size of the islands is typically of the order of tens of nanometers. Due to the Coulomb blockade effect, only one excess electron can be on an island at any time. This can be explained as follows. Due to their size, the islands have a very small capacitance and thus, an elementary charge $e$ on one islands induces a large potential. Therefore, a second electron is prevented to jump on the island before the first electron reached the neighbouring island (or the drain). We now apply a periodic pulse on each gate $U_1,\ldots,U_4$; the pulse on each gate $U_i$ is slightly retarded from the previous one $U_{i-1}$, such that these form a travelling wave which pushes the electron from the source to the drain.

Another essential role of the SEP is to synchronise different qubits (i.e. different branches of the calculation). By adjusting the timing between the gate pulses $U_1,\ldots,U_4$ we can make two electrons from different qubits to arrive simultaneously at the interaction region (usually a 2-qubit gate).

~\\
{\bf (ii) Measurement}\\
At the end of the calculation we need to measure the state of each qubit. This is done with a single electron transistor (SET), which is sensitive to sub-electron charges \cite{set}. A SET also uses the Coulomb blockade effect. If the source-drain voltage is just above the Coulomb threshold, the source-drain current is very sensitive to the gate voltage. Thus, a SET is an ideal tool to perform measurements of the single electron qubit.

~\\
{\bf (iii) Quantum Gates}\\
It is by now a classical result that any quputer can be build using only 1- and 2-qubit gates \cite{gates}. Our final task is then to show how to implement these gates in the proposed model.

~\\
{\bf Single qubit gates}\\
The only single qubit gates we need are phase shifters and beam splitters (equivalent to a Hadamard gate).

A phase shifter can be implemented with a quantum dot (QD). Using an Aharonov-Bohm ring with a QD embedded in one arm, several authors \cite{yacoby,schuster} have demonstrated experimentally that the transport through a quantum dot QD is coherent and that we can induce an arbitrary phase shift $\phi\in (0,\pi)$ by biasing the quantum dot gate with an adjustable voltage. Thus, by simply using a quantum dot biased by an appropriate gate voltage, we can induce the desired phase shift on one qubit.

A second single qubit gate is the Hadamard gate, which can be implemented using a beam splitter. One way is to use a symmetric tunnelling junction which makes an angle of $45^\circ$ with the direction of the electron. An electron wave incident on a tunnelling barrier will split into a reflected and a transmitted component and by adjusting the geometry of the barrier (height, width) we can have a symmetric beam splitter (ie the transmitted and reflected probabilities are equal).

However, a better (and somehow equivalent) way to implement an electronic beam splitter is to use an {\em electron waveguide coupler} \cite{alamo}-\cite{nl_coupler}. An electron waveguide coupler is formed from two parallel waveguides which are brought to an interaction region of length $L_c$.
\putfig{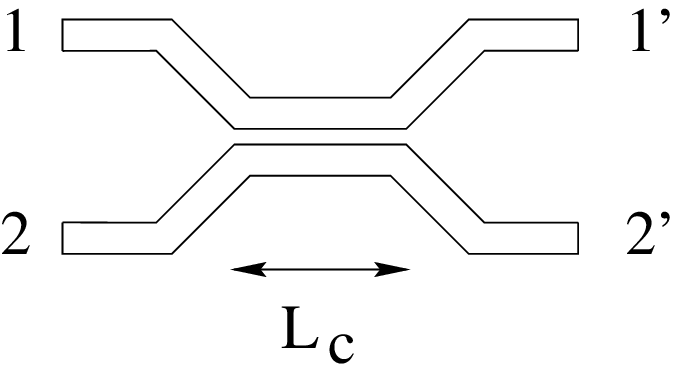}{3}
We only send an electron at input 1, leaving input 2 free. As the electron propagates along the waveguide, it oscillates back and forth between the two (due to the evanescent coupling between the waveguides). After a transfer length $L_t$, the electron injected initially in the upper waveguide is totally transferred in the lower one. In order to use this device as a symmetric beam splitter, we chose the coupling length of the gate to be half of the transfer length $L_c=L_t/2$. 

A typical transfer length for the complete electron transfer from one waveguide to the other is $0.28\mu$m \cite{tsukada}; therefore, a beam splitter can be made as small as $0.14\mu$m.

~\\
{\bf A 2-qubit gate}\\
The final step is to implement a two-qubit gate. A standard result is that almost any two-qubit gate is universal. Usually this gate is taken to be the {\sf CNOT}, mainly due to simplicity. In our case we prefer to have a conservative gate, ie a gate for which the number of {\em ones} at the input is the same as the number of {\em ones} at the output. For this we use a Coulomb coupler as a two-qubit gate.

The Coulomb coupler (CC) is described by the Hamiltonian \cite{c_coupler}:
\be
H= \hbar \chi N_A N_C
\ee
where $\chi$ is the coupling constant and $N_A$, $N_C$ are the particle number operators for the two qubits:

\be
N_A= a^\dagger a \ \ \ ,\ \ N_C= c^\dagger c
\ee

If $t$ is the interaction time, the effect of the gate on the two fields is:
\[ a \ra a'=a \ e^{-i\chi t N_C}  \ \ \ ,\ \ c \ra c'=c \ e^{-i\chi t N_A} \]
and thus the two electrons give each other a mutual phase modulation proportional to the particle number in each field. Since in our case $N_i=0$ or 1, the action of the gate can be written as

\[ \ket{00}\ra\ket{00} \ \ \ ,\ \ \ket{01}\ra \ket{01} \]
\be
\ket{10}\ra\ket{10} \ \ \ ,\ \ \ket{11}\ra e^{-2i\chi t}\ket{11}
\ee

The phase induced on each electron $\phi=\chi t$ is proportional to the coupling constant and to the interaction time, and hence to the gate length. By making the coupling $\chi$ sufficiently strong, the length of the gate can be made in principle less than $1\mu$m.

\begin{figure}
\putfig{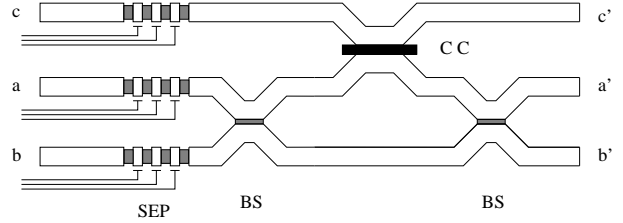}{8}
\caption{An electronic version of the quantum optical Fredkin gate. The beam splitters BS are electron waveguide couplers and the Coulomb coupler CC is used to induce a mutual phase modulation of the two qubits $a$ and $c$ (Kerr effect).}
\label{fig1}
\end{figure}

Combining two beam splitters and a Coulomb coupler, we can implement the electronic version of the optical quantum Fredkin gate proposed by Milburn \cite{milburn} (see Fig.\ref{fig1}).

We now discuss the advantages of a single electron implementation of a quputer over a single photon one. An important difference between the two is the coupling strength. The photon coupling is weaker than the electron coupling and therefore photons have a longer coherence time than electrons. However, due to this weak coupling, it is much more difficult to construct a 2 qubit gate which operates at single photon level. For example, in order to implement the optical quantum Fredkin gate \cite{milburn}, we need huge third-order susceptibilities $\chi^{(3)}$. On the other hand, it is much easier to construct a 2-qubit gate using single electrons than single photons. Also, it is much easier to detect and prepare single electron states than single photon states. Another important advantage is the large array of existing technologies in the field of electronics, compared to that of photonics.

There is a certain similarity between an electronic waveguide coupler and a Coulomb coupler. Although they operate differently (one is a single qubit and the other is a two qubit gate) and although they have different Hamiltonians, it is interesting to speculate that one could design a {\em single} gate which is programmable. Thus, by adjusting some external parameters (e.g. an external electric field or a potential barrier), this programmable gate can function in turn as a Coulomb coupler or as a beam splitter.

\section{Conclusions}

In this article we proposed a new model for a solid state quputer. We use ballistic single electrons as {\em flying qubits} in 1D nanowires. A single electron pump SEP is used to prepare the initial states of each qubit and also to synchronise different branches of the calculation. At the end of the calculation a single electron transistor SET measures the state of each qubit. Three types of gates are modelled. As single electron gates, we employ phase shifters and beam splitters (Hadamard gate). Phase shifters are simply implemented using quantum dots biased with the appropriate voltage. An electron waveguide coupler is equivalent to a beam splitter. As a two-qubit gate we use a Coulomb coupler which is the electronic analogue of the optical Kerr gate: if both electrons are present at the input, they give each other a phase modulation; otherwise, nothing happens.

To summarise, in the proposed model we
~\\
- use ballistic single electrons as {\em flying qubits} in 1D nanowires;\\
- use a single electron pump (SEP) to prepare the initial state of each qubit and to synchronise different branches of the calculation;\\
- use a single electron transistor (SET) to measure the final state of each qubit;\\
- use a quantum dot biased with an appropriate voltage as a phase shifter;\\
- use an electron waveguide coupler as a beam splitter;\\
- use a Coulomb coupler as a 2-qubit gate.

This model can be easily implemented using the present available technology. If a typical gate size is less than $1\mu$m, then for a phase coherence length of $30\mu$m several steps (tens) of a quantum computation can be executed before the system decoheres.

~\\
{\bf Acknowledgements}\\
We are grateful to Sandu Popescu and Mark Baxendale for useful discussions.

\end{document}